\documentclass[
  aps,
  prresearch,
  twocolumn,
  superscriptaddress,
  longbibliography
]{revtex4-2}
\usepackage{graphicx}
\usepackage{dcolumn}
\usepackage{bm}
\usepackage[version=4]{mhchem} 

\begin{document}

\title{Tuning cholesteric cellulose nanocrystal self-assembly in spherical confinement via salt and sonication}

\author{Diogo Vieira Saraiva}
\author{Anne Meike Hogeweg}
\author{Lisa Tran}
\email{l.tran@uu.nl}

\affiliation{
Soft Condensed Matter and Biophysics,
Debye Institute for Nanomaterials Science,
Utrecht University,
3584 CC Utrecht, The Netherlands
}


\begin{abstract}
Cellulose nanocrystals (CNCs) self-assemble into cholesteric liquid crystals that produce structural color upon solvent removal. Although most studies examine this process in planar films, confinement within micron-sized water-in-oil droplets provides a powerful platform for resolving self-assembly dynamics in real time. Here, we investigate how two common pitch-tuning strategies, sodium chloride addition and tip sonication, govern the kinetics and structure of CNC self-assembly under spherical confinement. Polarized optical microscopy timelapses capture the evolution from isotropic suspension through tactoid nucleation and annealing to kinetic arrest and final buckling. Consistent with prior work, pitch–concentration analysis reveals a universal post-arrest regime governed by droplet shrinkage. Beyond this established behavior, we identify a pre-arrest regime in which pitch decreases rapidly and whose kinetics accelerate with increasing salt concentration and sonication dose. These parameters shift the onset of cholesteric order and gelation, thereby tuning the concentration window for tactoid coalescence. Together, these results establish droplet confinement as a quantitative platform for probing out-of-equilibrium CNC self-assembly and for kinetically programming structurally colored soft materials.
\end{abstract}

\maketitle


\section{Introduction}
Cellulose nanocrystals (CNCs) dispersed in water spontaneously self-assemble into a cholesteric liquid crystalline phase once their concentration exceeds a critical volume fraction ($\phi$). This helical organization can be kinetically preserved during solvent removal, yielding solid material that has structural color. Most studies of this process employ planar films cast in petri dishes, where evaporation proceeds primarily through the water–air interface. As water is removed, the suspension experiences a uniaxial increase in concentration, driving the formation, coarsening, and eventual arrest of cholesteric order. Full evaporation produces a solid whose pitch is sufficiently compressed to reflect visible light, in accordance with Bragg’s law \cite{de_vries_rotatory_1951}.

As evaporation concentrates CNC suspensions, the system transitions from an initially isotropic fluid to discrete cholesteric tactoids, which grow and coalesce before undergoing kinetic arrest. After arrest, further water loss compresses the helical layers, reducing the pitch to sub-micron scales. Several approaches have been developed to tune the pitch of the final structure, including the addition of sodium chloride (\ce{NaCl}) \cite{araki_effect_2001, hirai_phase_2009}, which blue-shifts color by reducing the pitch, and tip-sonication, which red-shifts color by fragmenting chiral aggregates \cite{beck_controlling_2011, parton_chiral_2022, chiappini_modeling_2022}. Salt screens electrostatic repulsion by reducing the Debye length, while sonication alters particle morphology and bundle content, both of which modulate chiral interactions that govern self-assembly \cite{schutz_equilibrium_2020, frka-petesic_structural_2023}.

Although extensively characterized in planar films, these geometries complicate precise determination of the concentration at which kinetic arrest occurs because compression is nonuniform and uniaxial. Recent studies demonstrated that confining CNC suspensions within water-in-oil droplets produces markedly different assembly pathways \cite{parker_hierarchical_2016, parker_cellulose_2022, wang_polymer_2016, li_colloidal_2016}. In droplets, evaporation proceeds via water diffusion through the surrounding oil phase, generating radially symmetric concentration gradients. Combined with planar anchoring of CNCs at the droplet interface, this geometry promotes concentric cholesteric ordering and uniform volumetric compression, enabling more accurate identification of the concentration at kinetic arrest \cite{parker_hierarchical_2016}. Moreover, droplets permit direct, real-time visualization of pitch evolution in three dimensions, providing access to the out-of-equilibrium self-assembly dynamics preceding arrest. Despite these advantages, quantitative measurements of how common pitch-tuning parameters, such as salt concentration and sonication, influence self-assembly dynamics and shift the onset of kinetic arrest remain largely unexplored. 

Here, we extend our prior study of planar film geometries \cite{saraiva_controlling_2025} by examining the same CNC suspensions and pitch-tuning parameters under droplet confinement. The spherical confinement enables quantitative determination of how the concentration at kinetic arrest depends on salt concentration (\ce[NaCl]) and sonication dose, $u_s$. Using real-time polarized optical microscopy, we track the entire assembly trajectory, from the isotropic state to buckling of the final gel, while simultaneously quantifying pitch evolution. Consistent with prior observations, pitch–concentration scaling follows a universal post-arrest regime governed by droplet shrinkage \cite{parker_hierarchical_2016}. Building on this established behavior, we reveal a distinct pre-arrest regime where the pitch evolution under confinement is strongly modulated by both [\ce{NaCl}] and $u_s$. These two parameters influence the assembly pathway through different mechanisms: increasing [\ce{NaCl}] delays tactoid nucleation and promotes earlier gelation via enhanced electrostatic screening, thereby narrowing the tactoid-annealing window, whereas increasing $u_s$ weakens chiral interactions and shifts both the onset of cholesteric order and kinetic arrest to higher concentrations. Together, these results establish droplet confinement as a powerful, out-of-equilibrium platform for dissecting how interparticle forces and particle morphology govern liquid-crystalline self-assembly, and suggest a broadly applicable framework for probing kinetic arrest, phase evolution, and hierarchical ordering in confined colloidal and liquid-crystalline systems.

\section{Experimental}
\subsection{CNC synthesis}
Cellulose nanocrystals were synthesized by acid hydrolysis of Whatman no.1 cellulose filter paper (35 g). The filter paper was finely shredded using a spice grinder (Waring WSG60K). A 490 mL solution of 64 wt.-\% sulfuric acid (Sigma-Aldrich, 95-98\%) was prepared and heated to 66 $^\circ$C. The filter paper was introduced into the acid solution and stirred vigorously with a mechanical stirrer for 30 minutes to facilitate hydrolysis. The reaction was then quenched by adding five liters of deionized (DI) water. The resultant suspension was allowed to settle overnight, after which the clear supernatant was decanted. The sedimented material was centrifuged in 20-minute cycles at 15000 RCF until it formed a pellet. Further centrifugation cycles were conducted with the addition of DI water until the CNCs were suspended in the supernatant. The supernatant was collected and the residual pellet was discarded. The suspension was then dialyzed for two weeks in DI water (renewed daily) using MWCO 12-14 kDa dialysis membranes (Spectrum™ Spectra/Por™ 4). This process yielded a 1.1 wt.-\% CNC suspension with a total volume of approximately 900 mL.

\subsection{Preparation of CNC suspensions}
The stock CNC suspension was divided and diluted into separate suspensions (25 mL, [CNC] $=$ 4.35 wt.\%) in 50 mL centrifuge tubes (Corning™ Falcon™). These suspensions were tip sonicated using a Hielscher UP200St ultrasonic processor with a 7 mm titanium tip for various time intervals (corresponding to 0 $\leq$ $u_s$ $\leq$ 1440 J per mL of CNC suspension) under the following conditions: 30 W, 70:30 ON:OFF cycles, with the tip submerged halfway inside the sample. The suspensions were immersed in an ice bath during sonication to prevent sample heating and potential desulfation.

Following sonication, each of the suspensions was divided and diluted (with DI water and a 0.2 M aqueous \ce{NaCl} solution) into separate suspensions ([CNC] $=$ 3.5 wt.\%) with different \ce{NaCl} concentrations (0 $\leq$ [\ce{NaCl}] $\leq$ 450 mmol of \ce{NaCl} per kilogram of CNC). The CNC suspensions that underwent a combination of sonication dose ($u_s$) and salt addition ([\ce{NaCl}]) will be referenced in the convention of ["x" J/mL, "y" mmol/kg] for "x" J/mL and "y" mmol/kg. For example, [120 J/mL, 150 mmol/kg] refers to a CNC suspension with $u_s$ $=$ 120 J/mL and [\ce{NaCl}] $=$ 150 mmol/kg.

\subsection{Surface treatment of glass slides and vials}
All glass slides and vials were cleaned in a base bath for one hour, thoroughly rinsed with deionized water, dried under a nitrogen stream, and placed in an oven at 120 $^\circ$C overnight to eliminate residual moisture. A 0.2 M solution of octadecyltrichlorosilane (OTS, Sigma-Aldrich, 97\%) in anhydrous toluene (VWR, 99.8\%) was prepared. The glassware was immersed in the OTS solution for 90 minutes, followed by rinsing with anhydrous toluene. Finally, the treated glassware was dried overnight at 120 $^\circ$C.

\subsection{CNC droplet emulsions}
We prepared 2 mL of a solution of hexadecane (purchased from Thermo Fischer Scientific, 99\%) with 0.006 wt.-\% Span-80 (purchased from Sigma-Aldrich, $\geq$ 60\%). 25 {\textmu}L of the 3.5 wt.-\% CNC suspension was added. This emulsion was then manually shaken twice to form water droplets. 100 {\textmu}L of this emulsion was pipetted onto the OTS-coated glass slide immediately after shaking. All timelapse videos started within two minutes of casting.

\subsection{Polarized optical microscopy video timelapse}
The CNC droplets were imaged between crossed polarizers in transmission on a Nikon Ti-E inverted microscope equipped with a T-P2 DIC polarizer module and an analyzer. Images were recorded with a Nikon DS-Ri2 CMOS camera. The framerate of the timelapses was set to 1 frame per minute and the timelapses were performed under ambient conditions (20 $^\circ$C, relative humidity of 40-50\%). The duration of the timelapses varied between 4 and 30 hours, depending on the volume and parameters of the droplet.

\subsection{Volume fraction and pitch size measurements}
The pitch evolution graphs were plotted by calculating the average CNC volume fraction and average cholesteric pitch size every two frames using ImageJ 1.53t software. In each of the measured frames, two diameter measurements and five half-pitch size measurements were measured by hand and averaged. Knowing the initial CNC volume fraction ($\phi$ $\approx$ 2.2 vol.\%) and spherical diameter, the CNC volume fraction of the droplets was calculated for each frame. Each pitch value consisted of five half-pitch measurements within a chosen region of the droplet. The region used for the measurements was kept fixed throughout the timelapse. Only tactoids with clearpitch layers were measured in each frame.

\section{Results and discussion}
\subsection{Polarized optical microscopy timelapses} 

\begin{figure*}[t]
\centering
  \includegraphics[height=10cm]{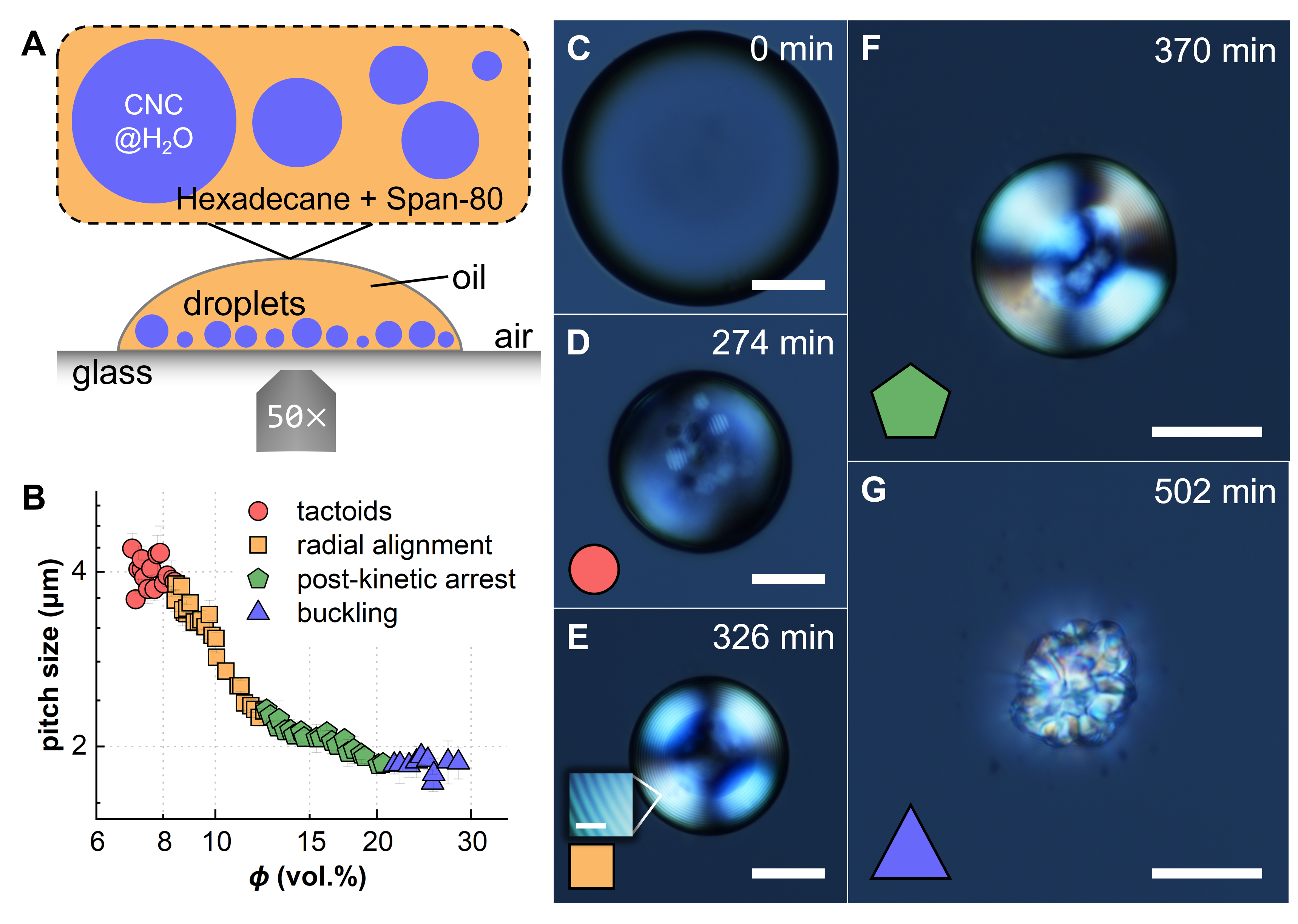}
  \caption{(A) Schematic of CNC self-assembly within a water droplet. The emulsion droplets consist of CNCs in water (blue), dispersed in a hexadecane/Span-80 continuous phase (yellow). When deposited onto a fluorophobic glass slide, the CNC droplets rest on the surface while largely maintaining their spherical geometry. (B) Log-log plot showing the evolution of cholesteric pitch size (y-axis) as a function of CNC volume concentration ($\phi$, x-axis) for a [8 J/mL, 150 mmol/kg] CNC droplet. The self-assembly pathway is divided into four stages, color-coded on the plot: tactoid stage (red circles, D), radial alignment stage (yellow squares, E), post-kinetic-arrest stage (green pentagons, F), and buckling stage (blue triangles, G). (C)-(G) Crossed-polarized micrographs illustrating these stages for the same sample: (C) isotropic phase, showing no birefringence; (D) nucleation and growth of tactoids; (E) coalescence of tactoids and anchoring of cholesteric layers to the droplet perimeter. The inset in (E) provides a clearer view of the radially aligned cholesteric pitch; (F) onset of kinetic arrest; (G) fully buckled droplet after complete water diffusion. Scale bars for (C)-(G): 50 {\textmu}m. Scale bar for inset of (E): 5 {\textmu}m.}
  \label{fgr:main1}
\end{figure*}

We measured how the cholesteric pitch size changes during water diffusion into the surrounding hexadecane, depending on the estimated volume concentration of CNCs. Figure \ref{fgr:main1} presents key frames of an example timelapse video (Supplementary Information Video 2) for a [8 J/mL, 150 mmol/kg] sample. Figures \ref{fgr:main1}C to \ref{fgr:main1}G depict representative polarized optical microscopy snapshots, corresponding to successive stages of cholesteric assembly. Initially, the CNC suspension is isotropic and the droplet appears transparent under crossed polarizers. After 274 minutes, the first tactoids become visible. By 326 minutes, the tactoids have coalesced and the cholesteric structure has aligned predominantly radially at the water-oil interface, clearly shown in the inset of Figure \ref{fgr:main1}. At 370 minutes, the structure ceases to grow or move appreciably, indicating the onset of kinetic arrest. From this moment until the end of the timelapse (502 minutes), residual water is removed and the structure buckles. Figure \ref{fgr:main1}B  shows the corresponding pitch evolution, with points color-coded according to these visually defined stages: tactoid formation (red circles), radial alignment (yellow squares), post-kinetic-arrest (green pentagons), and buckling (blue triangles).

The pitch evolution graph in Figure \ref{fgr:main1}B resembles trends reported previously\cite{parker_hierarchical_2016, parker_cellulose_2022}. Across the full assembly process, the pitch size decreases with increasing CNC concentration, consistent with tighter particle packings strengthening the cholesteric twisting power and thus reducing the pitch. A distinct change in scaling appears at $\phi$ $\approx$ 13 vol.\%, marking the onset of kinetic arrest. 

To interpret how the pitch decreases prior to kinetic arrest, we consider each assembly stage. Before kinetic arrest, during the tactoid stage (\ref{fgr:main1}B, red circles), the pitch decreases slowly and irregularly. As the process transitions into the radial-alignment stage (\ref{fgr:main1}B, yellow squares), the pitch decreases more rapidly and approximately linearly. At this point, tactoids have largely coalesced at the water-oil interface into a single cholesteric domain. Anchoring of the cholesteric layers to the shrinking droplet boundary forces the pitch to reduce proportionally to the droplet volume.

At and beyond the trend shift in the pitch evolution (\ref{fgr:main1}B, green pentagons), the system enters the kinetically arrested regime. This transition has been similarly identified in prior studies \cite{parker_hierarchical_2016, parker_cellulose_2022}. Here, the rate of pitch decreases becomes markedly slower. Kinetic arrest results from concurrent increases in ionic content and CNC volume fraction during solvent diffusion, which induce gelation \cite{schutz_equilibrium_2020, honorato-rios_fractionation_2018}. Once the CNCs gelate, the cholesteric structure becomes spatially frozen, and the evolution of the pitch size results from compression of the gelled structure with continued water evaporation. For all samples in this study, the kinetic arrest point is identified from both the trend shift in the pitch size and from direct visual cues in the timelapse videos (Supplementary Information Videos 1-5).

Finally, after the cholesteric phase transitions into a gel, continued water evaporation from the droplet generates radial mechanical stress that ultimately produces buckling. During the buckling stage (\ref{fgr:main1}B, blue triangles), the measured pitch appears nearly stagnant because buckling produces distortions reminiscent of the Helfrich-Hurault elastic instability \cite{helfrich_electrohydrodynamic_1971, hurault_static_1973, blanc_helfrich-hurault_2023}, in which undulation of the cholesteric layers allows the material to accommodate volume changes while maintaining the layer spacing. Pitch measurements in these frames are particularly error-prone, due to both the buckling deformations and the proximity to the optical resolution limit. Consequently, pitch values were not measured for frames exhibiting substantial buckling.

\begin{figure}[t]
\centering
  \includegraphics[height=9cm]{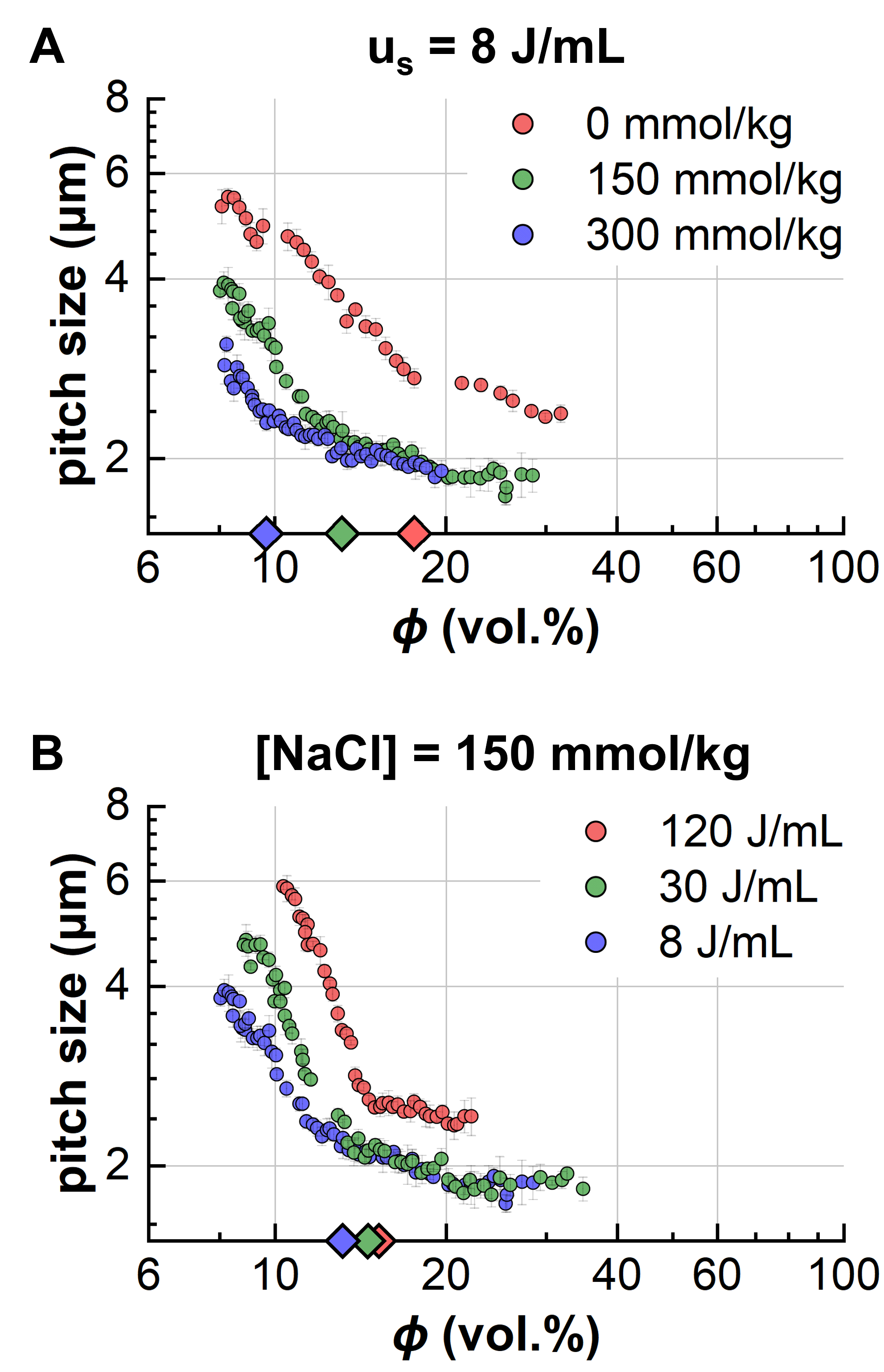}
  \caption{Evolution of the cholesteric pitch size (y-axis) as the CNC volume fraction increases (x-axis). Colored diamonds on the x-axis mark the volume fraction at which each sample undergoes kinetic arrest. (A) Three samples with increasing [\ce{NaCl}] values at a fixed $u_s= 8$ J/mL; (B) Three samples with increasing $u_s$ values at a fixed [\ce{NaCl}] of 150 mmol/kg.}
  \label{fgr:main2}
\end{figure}
We measured the variation in pitch size for samples prepared with five different combinations of salt concentration and sonication dose. Figure \ref{fgr:main2}A isolates the effect of electrolyte addition by fixing the sonication dose at $u_s=8$ J/mL and varying [\ce{NaCl}] from 0 to 300 mmol/kg. Figure \ref{fgr:main2}B, conversely, isolates the effect of sonication by fixing [\ce{NaCl}] at 150 mmol/kg instead and varying $u_s$ from 8 to 120 J/mL. Together, the graphs allow examination of how salt and sonication independently modulate the cholesteric self-assembly pathway.

Beginning with the influence of salt concentration (Figure \ref{fgr:main2}A), increasing [\ce{NaCl}] at fixed $u_s = 8$ J/mL reduces the cholesteric pitch size throughout the assembly process, consistent with previous work \cite{dong_effects_1996}. This pitch size reduction is particularly pronounced during the tactoid and radial alignment stages. Measurable pitch values in tactoids emerge at similar CNC volume fractions ($\phi$ $\approx$ 6-7 vol.\%) across all salt concentrations. Importantly, this onset of measurable pitch values should not be conflated with the equilibrium critical concentration for cholesteric ordering (the first timelapse frame of birefringence), which does increase with salt addition \cite{dong_effects_1996}. As indicated by the colored diamonds in Figure \ref{fgr:main2}A, the kinetic-arrest volume fraction decreases with increasing [\ce{NaCl}]. A shorter Debye length reduces interparticle repulsions, promoting aggregation and thus triggering kinetic arrest at lower concentrations.

Turning to the effect of sonication dose (Figure \ref{fgr:main2}B), increasing $u_s$ at fixed [\ce{NaCl}] = 150 mmol/kg produces the opposite trend: the overall pitch size increases during both the tactoid and radial-alignment stages, as expected from previous studies \cite{beck_controlling_2011}. For all samples and for any given volume fraction, higher $u_s$ corresponds to a larger pitch. The concentration at which tactoid formation begins also increases with sonication dose, consistent with the idea that bundle breakage reduces the population of chiral dopants needed for cholesteric ordering \cite{parton_chiral_2022}. Consequently, the onset of kinetic arrest shifts to higher concentrations as $u_s$ increases.

\subsection{Quantifying the rate of pitch reduction}
\begin{figure*}[t]
\centering
  \includegraphics[height=9cm]{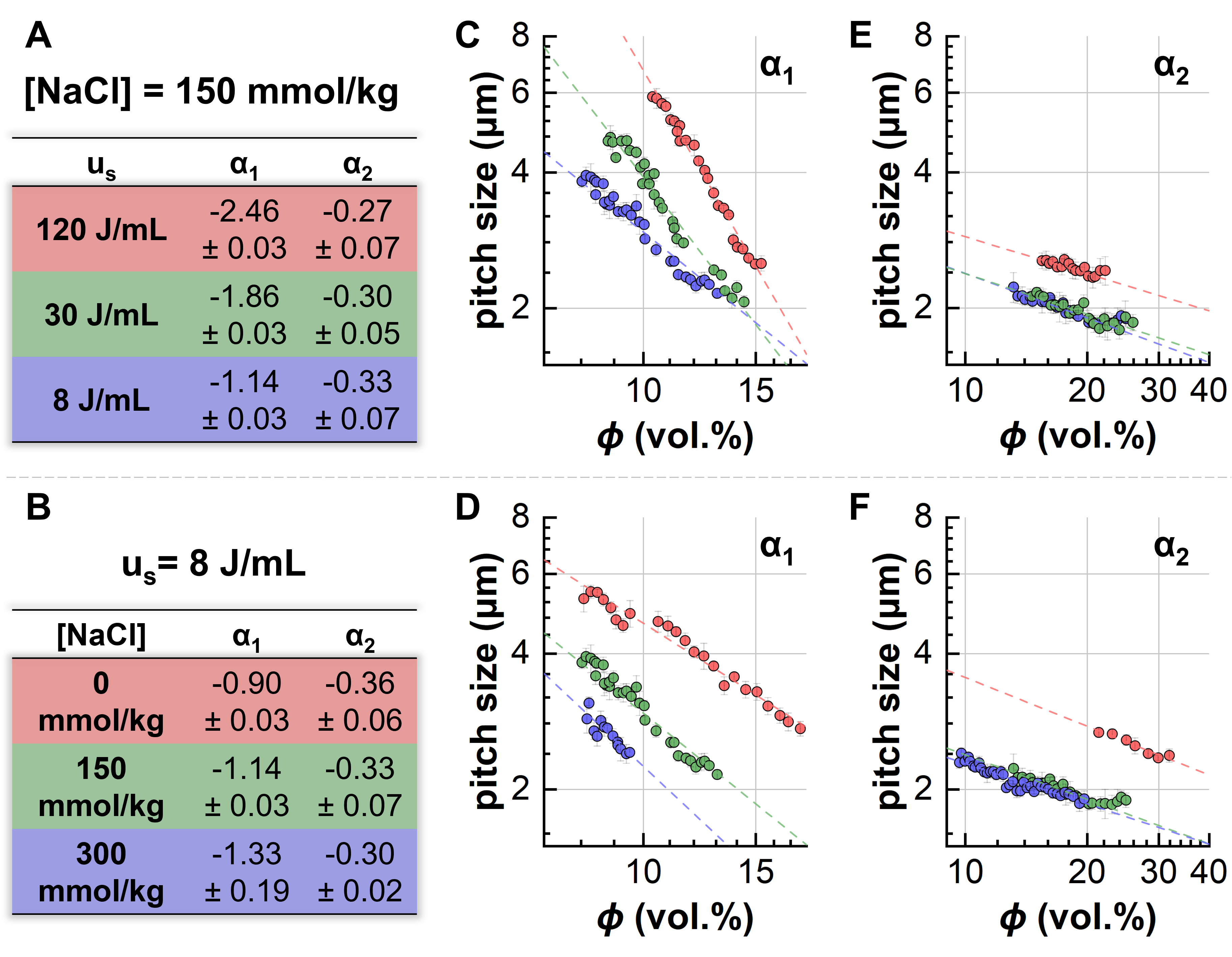}
  \caption{Values of ${\alpha_1}$ and ${\alpha_2}$ values for (A) samples of increasing $u_s$ at a fixed [\ce{NaCl}] = 150 mmol/kg, and (B) samples of increasing [\ce{NaCl}] at a fixed $u_s$ = 8 J/mL. (C), (D) Pitch-size data before kinetic arrest for increasing $u_s$ and increasing [\ce{NaCl}], respectively, with trend lines of the form $p = k\phi^{\alpha_1}$. (E), (F) Pitch-size data after kinetic arrest for the same sample sets, fitted with $p = k\phi^{\alpha_2}$.}
  \label{fgr:main3}
\end{figure*}

The pitch evolution graphs in Figure \ref{fgr:main2} were fitted with trend lines of the form $p = k\phi^{\alpha}$, with $\phi$ being the CNC volume fraction, evaluated separately before ($\alpha_{1}$) and after ($\alpha_{2}$) the onset of kinetic arrest. Figure \ref{fgr:main3} presents these fitted trends lines along with the extracted values of $\alpha_1$ and $\alpha_2$ for each sample.

The exponents $\alpha_1$ and $\alpha_2$ quantify the rate at which the pitch decreases with increasing CNC concentration. Because pitch always decreases as concentration increases, both exponents are negative; larger-magnitude (more negative) values correspond to faster pitch reduction. Two earlier studies by Parker et al. \cite{parker_hierarchical_2016, parker_cellulose_2022} analyzed pitch–concentration scaling in a similar droplet-based CNC system and reported $\alpha_1$ values of $-1$ and $-1.9$ for two CNC batches with different source materials and salt concentrations. These works helped establish the sensitivity of pre-arrest pitch reduction to the details of CNC morphology and solution conditions. Building on this foundation, our study extends the analysis by systematically varying both salt concentration and sonication dose for a single CNC source, enabling a direct comparison of how these commonly used tuning parameters modulate the early-stage cholesteric relaxation regime.

We report five distinct $\alpha_1$ values for the five analyzed samples that vary in salt concentration and sonication dose, given in Figure \ref{fgr:main3}. Across samples, $\alpha_1$ systematically decreases with either increasing salt concentration or increasing sonication dose. When [\ce{NaCl}] increases from 0 mmol/kg to 300 mmol/kg, $\alpha_1$ decreases from $-0.90$ to $-1.33$. Increasing $u_s$ from 8 J/mL to 120 J/mL lead to a more dramatic decrease in $\alpha_1$, from $-1.14$ to $-2.46$. In contrast, all $\alpha_2$ values cluster consistently around $-1/3$, aligning closely with previously reported values \cite{parker_hierarchical_2016, parker_cellulose_2022}. Complementary time-dependent pitch reduction rates (Supplementary Information Figure 1) further show that higher [\ce{NaCl}] reduces the pitch reduction rate, whereas increasing $u_s$ produces a more complex, non-monotonic response over the time-frame of the droplet evaporation.

To interpret these exponents, we first consider the meaning of $\alpha_2$. All samples exhibit $\alpha_2 \approx -1/3$, identical to earlier reports. This exponent originates from the geometrical relation between pitch ($p$), droplet volume ($V$), and CNC concentration ($C$): for a radially aligned cholesteric in a spherical geometry, $p \propto V^{1/3} \propto C^{-1/3}$. Such scaling is expected only when the pitch reduction is governed exclusively by droplet shrinkage. After kinetic arrest, the cholesteric material forms an attractive gel, so layer compression occurs solely through water loss rather than internal rearrangement.

In contrast, the $\alpha_1$ values are substantially more negative than $\alpha_2$, indicating that pitch reduces far more rapidly before kinetic arrest. Equilibrium phase diagrams do not capture this fast, out-of-equilibrium compression (at kinetic arrest the system becomes trapped in a non-equilibrium state locked by gelation and geometric constraints \cite{honorato-rios_equilibrium_2016, gparton_angle-resolved_2024}), underscoring that $\alpha_1$ reflects a dynamic, transient regime absent in equilibrium systems. In this early stage, cholesteric domains can grow, flow, and coalesce, allowing the structure to optimize packing. Such optimization increases the cholesteric twisting power, producing a tighter helix and reducing the pitch \cite{tran_fabrication_2018}. This mechanism helps explain why both added salt and increased sonication accelerate pitch reduction with increasing CNC concentration. Each reduces the effective CNC size: salt by thinning the electrostatic double layer, and sonication by fragmenting aggregates and bundles. Smaller average particle sizes enable more efficient packing, thereby increasing the rate of pitch reduction. We confirmed the reduction in average particle size with increased sonication dose through AFM data presented in our prior work on planar films \cite{saraiva_controlling_2025}.

The distinct behaviors of $\alpha_1$ and $\alpha_2$ thus reflect two fundamentally different regimes of droplet evolution: an initial dynamic, cholesteric relaxation followed by a later diffusion-limited stage constrained by gelation. To determine whether these observed pitch-reduction dynamics are governed primarily by evaporation kinetics or by the physicochemical modifications induced by salt and sonication, it is necessary to examine the rate of water loss from the droplets.

\subsection{Analysis of water diffusion rate}
To quantify the water diffusion rate across samples, we applied the classical $D^2$ law of droplet evaporation \cite{dalla_barba_revisiting_2021}. The $D^2$ law states that, for a diffusion-limited evaporation of a spherical droplet, the square of the droplet radius decreases linearly with time:
\[
r^2(t) = r^2_0 - Kt
\]
where $r_0$ is the initial radius and $K$ is the evaporation constant. Although this model is normally used for droplets evaporating in air, the same physical arguments extend to evaporation of a water droplet in an oil medium \cite{krishan_evaporation_2024}, provided that the droplet remains approximately spherical and evaporation is governed by diffusion of water into the surrounding oil.

\begin{figure}[t]
\centering
  \includegraphics[height=5.5cm]{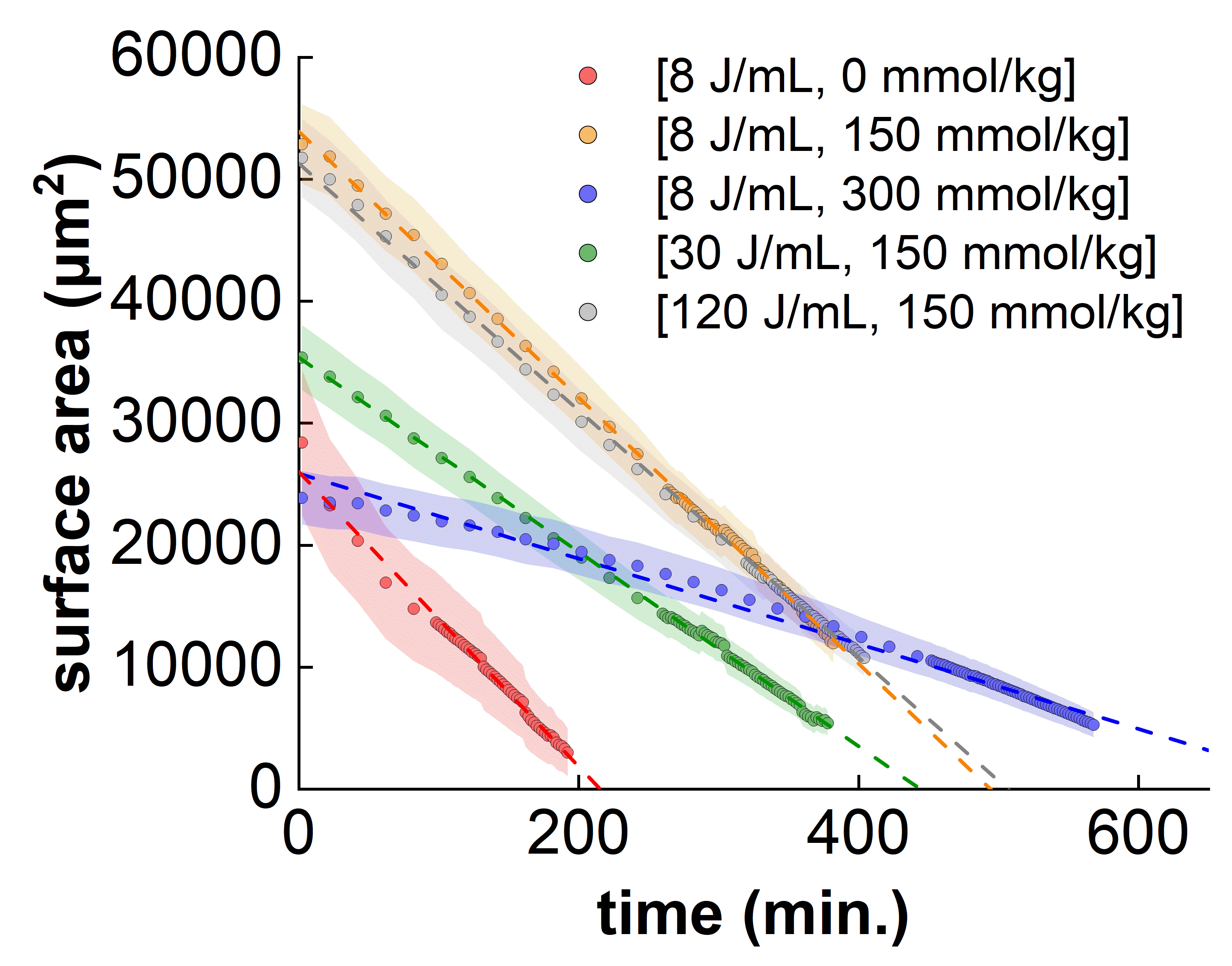}
  \caption{Droplet surface area plotted over time for droplets with varying salt and sonication levels. Linear fits (dashed lines) follow the $D^2$ law. Shaded regions indicate the propagated uncertainty from a $\pm$5 {\textmu}m error in measured droplet diameter.}
  \label{fgr:main4}
\end{figure}

In Figure \ref{fgr:main4}, the droplet surface area ($A \propto r^2$) is plotted as a function of time for each sample, and a linear fit of the form $A_0 - Kt$ was applied. The slope of the fit, $K$, represents the evaporation rate, while $A_0$ represents the surface area at $t = 0$. The resulting evaporation rate constants ($K$) and their surface-area-normalized forms $K/A_0$ are summarized in Table \ref{tbl:evaporation_slopes}.

\begin{table}[t]
\small
  \caption{\ Evaporation constants ($K$) and surface-area-normalized evaporation rates ($K/A_0$) for all measured droplets}
  \label{tbl:evaporation_slopes}
  \begin{tabular*}{0.48\textwidth}{@{\extracolsep{\fill}}lll}
    \hline
    Sample & $K\,(\mu\text{m}^2.\text{s}^{-1})$ & $K/A_0\,(\text{s}^{-1})$ \\
    \hline
    \text{\,[8 J/mL, 0 mmol/kg]} & $-241 \pm 4$ & $(-9.0 \pm 0.1)\times 10^{-3}$ \\
    \text{\,[8 J/mL, 150 mmol/kg]} & $-218 \pm 1$ & $(-4.00 \pm 0.02)\times 10^{-3}$ \\
    \text{\,[8 J/mL, 300 mmol/kg]} & $-70 \pm 1$ & $(-3.0 \pm 0.1)\times 10^{-3}$ \\
    \text{\,[30 J/mL, 150 mmol/kg]} & $-159 \pm 1$ & $(-5.00 \pm 0.03)\times 10^{-3}$ \\
    \text{\,[120 J/mL, 150 mmol/kg]} & $-202 \pm 1$ & $(-4.00 \pm 0.02)\times 10^{-3}$ \\
    \hline
  \end{tabular*}
\end{table}

When comparing the unnormalized evaporation constants $K$, no clear trend emerges between droplets containing 0 and 150 mmol/kg [\ce{\ce{NaCl}}], and increasing sonication dose likewise does not yield a discernible trend in $K$. This lack of a trend in $K$ can be attributed to the varying initial droplet sizes across samples. Because each droplet begins with a different initial surface area, the evaporation constants $K$ can be normalized by $A_0$ to compare droplets with differing initial sizes, shown in the rightmost column of Table~\ref{tbl:evaporation_slopes}. This normalization highlights that evaporation becomes progressively slower with increasing salt concentration, while sonication dose induces no consistent change in the evaporation. Notably, the droplet with [8 J/mL, 300 mmol/kg] exhibits a markedly slower evaporation rate. A slower water-loss rate at higher salt concentration is expected \cite{mor_effect_2018}, and is consistent with the reduced $K/A_0$ values observed for across samples with increasing salt. Conversely, no systematic trend is observed with increasing $u_s$, indicating that sonication does not substantially influence water diffusion.

Comparing these evaporation rates with the corresponding $\alpha_1$ values in Figure \ref{fgr:main3} clarifies the origins of the observed pitch reduction before kinetic arrest. The strong dependence of $\alpha_1$ on $u_s$ cannot be attributed to evaporation kinetics, as $K$ remains essentially unchanged with increasing sonication dose. This confirms that the sonication-induced changes in $\alpha_1$ arise from morphological modifications to the CNCs themselves, rather than changes in water loss.

In contrast, increasing salt concentration simultaneously increases the pitch reduction rate and slows water diffusion. This coupling makes it more difficult to disentangle the contributions of evaporation kinetics and electrostatic screening. The time-dependent pitch evolution in Supplementary Information Figure 1 supports this interpretation: higher salt concentrations slow both water diffusion and pitch reduction, mirroring the trends observed in Figure \ref{fgr:main4}. Overall, salt simultaneously accelerates cholesteric relaxation and slows water evaporation, producing the observed combination of faster pitch reduction and slower water loss.

\subsection{Stages of assembly}
\begin{figure*}[t]
\centering
  \includegraphics[width=0.7\textwidth]{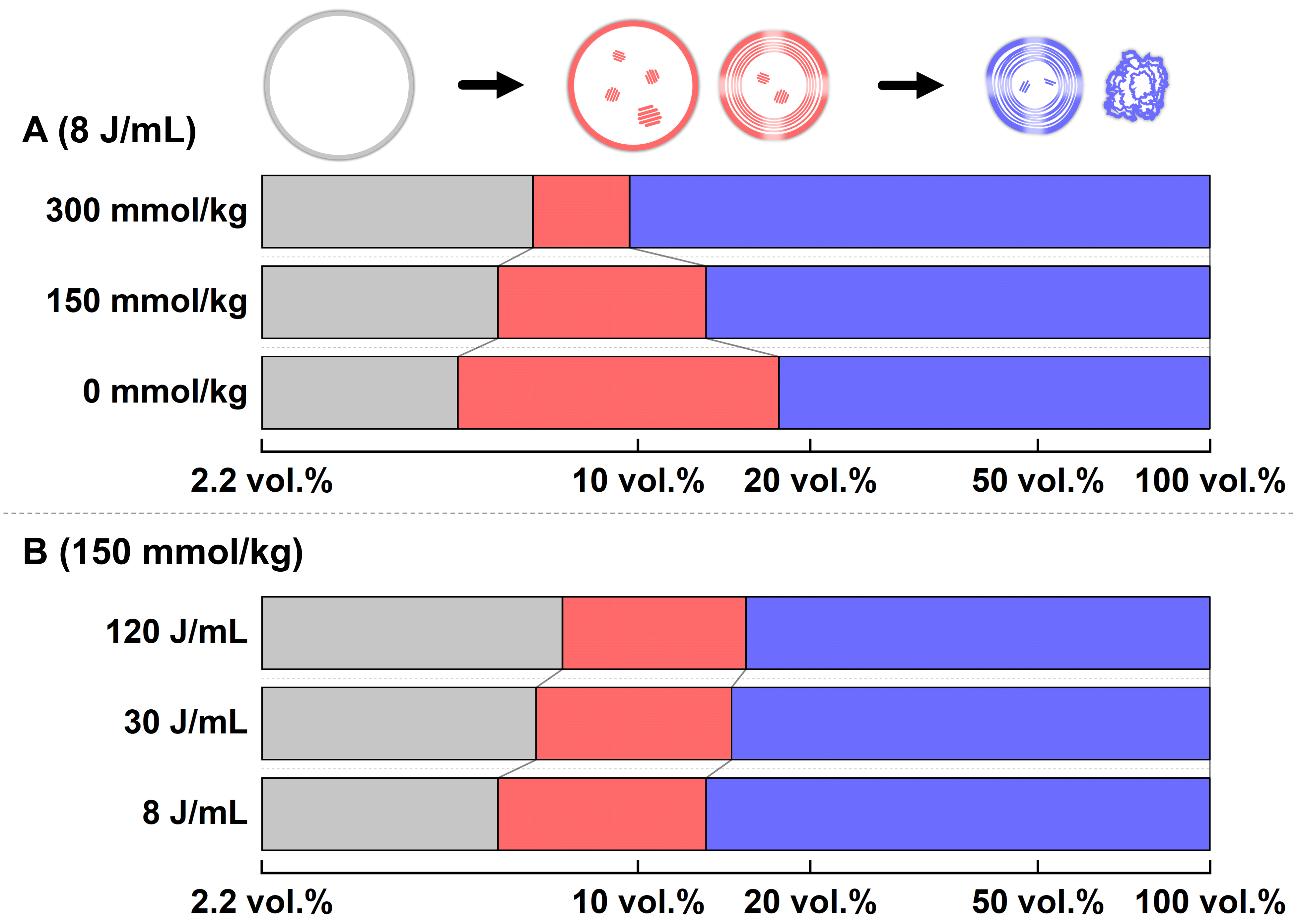}
  \caption{Progress bars showing the different stages of self-assembly. All samples begin at a CNC volume fraction ($\phi$) of 2.2 vol.\%. The isotropic phase is shown in gray; the tactoid nucleation and annealing stage in red; and the gelled, kinetically arrested stage in blue. Each progress bar ends at $\phi = 100$ vol.\%, corresponding to complete water loss.}
  \label{fgr:main5}
\end{figure*}

Each sample's timelapse video and pitch evolution graph provide a chronological map of the self-assembly stages that occur as water diffuses out of the droplet. By analyzing the videos frame-by-frame, we identified the transitions between the three key stages of CNC ordering: i) an initial isotropic stage, (ii) a tactoid nucleation and annealing stage, and (iii) a gelled, kinetically arrested stage. For each droplet, these stages are represented as progress bars in Figure \ref{fgr:main5}.

Figure \ref{fgr:main5} can be viewed as an out-of-equilibrium analogue to a phase diagram. Unlike conventional equilibrated phase diagrams, where phase boundaries depend solely on CNC volume fraction, here the transition occurs dynamically as $\phi$ increases continuously during evaporation. All samples begin in the isotropic stage at $\phi$ = 2.2 vol.\%, corresponding to the initial CNC concentration of 3.5 wt.\%, and the samples all end at $\phi$ = 100 vol.\%, after complete water diffusion from the droplet. The tactoid annealing stage begins when birefringence first becomes detectable and ends when kinetic arrest sets in. As earlier sections have shown, increasing [\ce{NaCl}] reduces the $\phi$ value required for kinetic arrest, whereas increasing $u_s$ shifts arrest to higher $\phi$.

In Figure \ref{fgr:main5}A, increasing \ce{NaCl} substantially narrows the tactoid annealing window. Salt delays the isotropic-to-cholesteric transition because electrostatic screening reduces the CNC effective volume, requiring a higher $\phi$ before tactoids nucleate. At the same time, stronger attractive interactions, driven by the reduced Debye length, induce kinetic arrest at progressively lower $\phi$. Together, these effects leave less time for tactoid coarsening and coalescence, which hinders the formation of large and well-ordered cholesteric domains. This behavior is especially striking in highly salted samples, such as [0 J/mL, 450 mmol/kg] (Supplementary Information Video 7), where tactoids gel within roughly twenty minutes of nucleation.

In contrast, Figure~\ref{fgr:main5}B shows that increasing sonication dose shifts both tactoid nucleation and kinetic arrest to slightly higher $\phi$. This trend aligns with previous reports that sonication-induced fragmentation reduces chiral strength \cite{beck_controlling_2011,gicquel_impact_2019}, increasing the particle volume fraction needed to satisfy Onsager-like packing that promotes cholesteric ordering. While some studies find minimal change in the cholesteric concentration with sonication \cite{parton_chiral_2022}, our observations indicate that changes to the CNC particle morphology from sonication weakens long-range helical interactions sufficiently to influence the onset of cholesteric order under confinement. 

Interestingly, increasing the sonication dose does not dramatically alter the duration of the intermediate tactoid-annealing stage. This behavior may arise from several possible mechanisms. First, tactoid growth and coalescence in droplets are strongly governed by evaporation-driven flows and geometric confinement, which may dominate over changes in particle-level chirality. Second, the spherical confinement may impose a relatively fixed coalescence pathway, effectively standardizing the timescale of tactoid annealing across samples with similar salt conditions. The precise reasons for this apparent insensitivity of the duration of tactoid annealing to sonication under fixed salt conditions remain unclear and warrant further investigation.

\section*{Conclusions}
We presented a confinement-based method to probe CNC self-assembly dynamics by tracking the evolution of cholesteric pitch in evaporating aqueous droplets surrounded by oil. Timelapse imaging enables precise identification of the isotropic stage, tactoid nucleation and coalescence, kinetic arrest, and final buckling. By quantifying how these transitions shift under varying salt concentration and sonication dose, we decoupled the effects of electrostatic screening, morphological modification, and evaporative compression on out-of-equilibrium cholesteric formation.

Salt and sonication influence the assembly pathway in distinct ways. Increasing [\ce{NaCl}] both accelerates pitch reduction and induces earlier kinetic arrest by enhancing interparticle attractions and reducing the effective particle volume. Sonication, in contrast, delays cholesteric formation and arrest by reducing the helical twisting power, while leaving the rate of water loss largely unaffected. The pre-arrest pitch-reduction rate ($\alpha_1$) increases with either parameter, reflecting a combination of concentration-driven pitch relaxation during cholesteric annealing and more efficient local packing of CNCs as electrostatic repulsion is reduced or bundles are fragmented. After arrest, all samples converge to the same scaling law ($\alpha_2 = -1/3$), indicating that late-stage pitch evolution is governed solely by volume shrinkage, consistent with prior observations \cite{parker_hierarchical_2016}.

Beyond mapping how CNCs assemble under confinement, this study highlights how surface anchoring and strong geometric confinement fundamentally shape the assembly pathway. Radial confinement in droplets imposes boundary conditions that favor concentric helicoidal ordering and promote the growth and annealing of tactoids, processes that can be hindered or entirely absent in planar film geometries where the system is less confined and compression is uniaxial. The spherical interface provides a boundary that stabilizes cholesteric domains, enhances tactoid coalescence, and supports ordering at lower shear and defect densities than in films. These confinement-induced advantages make droplets a uniquely sensitive platform for resolving transient assembly regimes, including the approach to kinetic arrest.

More broadly, the droplet-based method offers a framework for comparing equilibrium phase behavior with nonequilibrium assembly trajectories. Future work could leverage this platform to investigate the self-assembly kinetics of multicomponent systems. Ultimately, bridging the gap between dynamic assembly pathways and static phase diagrams will support more predictive control over CNC-based photonic materials.

\section*{Data availability}
The Supplementary Information (SI) document contains clarification on power law fit and error estimation, a brief analysis on the evolution of cholesteric pitch size in function of time, and videos of all droplet evaporation timelapses.

\section*{Acknowledgements}
The authors thank Dave van den Heuvel and Relinde van Dijk-Moes for helpful discussions and technical assistance. We thank Ivo Vermaire for help with glass functionalization. D.V.S. acknowledges financial support from the Department of Physics, Utrecht University. L.T. acknowledges support from the European Commission (Horizon-MSCA, Grant No. 892354) and the Dutch Research Council NWO ENW Veni grant (Project No. VI.Veni.212.028).


\bibliography{references} 
\bibliographystyle{rsc} 

\end{document}